\documentclass[runningheads]{llncs}

\usepackage{amsmath}
\usepackage{fontawesome}
\usepackage{multirow}
\usepackage[table,xcdraw]{xcolor}
\usepackage{graphicx}
\usepackage{textcomp}
\usepackage{subfig}
\usepackage[utf8x]{inputenc}
\usepackage{csquotes}
\usepackage{flexisym}
\newcommand{\ignore}[1]{}
\usepackage[hidelinks]{hyperref}
\usepackage{hyphenat}
\usepackage[ruled,vlined,linesnumbered]{algorithm2e}
\usepackage{amssymb}
\usepackage[bottom,multiple]{footmisc}
\usepackage{wrapfig}
\usepackage{nicefrac}
\usepackage{bm}
\usepackage{mathrsfs}
\usepackage[font=scriptsize]{caption}

\newcommand{\pactivities}{\mathcal{A}}

\newcommand{\pcases}{\mathcal{C}}

\newcommand{\ptimes}{\mathcal{T}}

\newcommand{\multiset}{\mathcal{B}}

\newcommand{\universe}{\mathcal{U}}

\newcommand{\eventlog}{L}
\newcommand{\tlkc}{\mathrm{TLKC}}

\newtheorem{exmp}{Example}

\DeclareMathSymbol{\mlq}{\mathord}{operators}{'134}
\DeclareMathSymbol{\mrq}{\mathord}{operators}{'42}

\usepackage{mathtools}

\begin{document}
\title{Towards Quantifying Privacy in Process Mining}
\titlerunning{Towards Quantifying Privacy in Process Mining}
%
%
\author{Majid Rafiei\orcidID{0000-0001-7161-6927}\textsuperscript{\faEnvelopeO} \and
	Wil M.P. van der Aalst\orcidID{0000-0002-0955-6940}}
\authorrunning{Majid Rafiei and Wil M.P. van der Aalst}
%
\institute{Chair of Process and Data Science, RWTH Aachen University, Aachen, Germany \\
 }
\maketitle              

\begin{abstract}
Process mining employs event logs to provide insights into the actual processes. Event logs are recorded by information systems and contain valuable information helping organizations to improve their processes. 
However, these data also include highly sensitive private information which is a major concern when applying process mining. 
Therefore, privacy preservation in process mining is growing in importance, and new techniques are being introduced. 
The effectiveness of the proposed privacy preservation techniques needs to be evaluated. It is important to measure both sensitive data protection and data utility preservation.
In this paper, we propose an approach to quantify the effectiveness of privacy preservation techniques. We introduce two measures for quantifying disclosure risks to evaluate the sensitive data protection aspect. Moreover, a measure is proposed to quantify data utility preservation for the main process mining activities. The proposed measures have been tested using various real-life event logs.


\end{abstract}
\section{Introduction}\label{sec:introduction}
Process mining bridges the gap between traditional model-based process analysis (e.g., simulation), and data-centric analysis (e.g., data mining) \cite{van2016process}. 
The three basic types of process mining are \textit{process discovery}, where the aim is to discover a process model capturing the behavior seen in an event log, \textit{conformance checking}, where the aim is to find commonalities and discrepancies between a process model and an event log, and \textit{process re-engineering} (\textit{enhancement}), where the idea is to extend or improve a process model using event logs.

An event log is a collection of events. Each event has the following mandatory attributes: a \textit{case identifier}, an \textit{activity name}, a \textit{timestamp}, and optional attributes such as \textit{resources} or \textit{costs}. 
In the human-centered processes, case identifiers refer to individuals. For example, in a patient treatment process, the case identifiers refer to the patients whose data are recorded. 
Moreover, other attributes may also refer to individuals, e.g., \textit{resources} often refer to persons performing activities. 
When event logs explicitly or implicitly include personal data, \textit{privacy concerns} arise which should be taken into account w.r.t. regulations such as the European General Data Protection Regulation (GDPR).


The \textit{privacy} and \textit{confidentiality} issues in process mining are recently receiving more attention and various techniques have been proposed to protect sensitive data.
Privacy preservation techniques often apply anonymization operations to modify the data in order to fulfill desired privacy requirements, yet, at the same time, they are supposed to preserve data utility. 
To evaluate the effectiveness of these techniques, their effects on \textit{sensitive data protection} and \textit{data utility preservation} need to be measured. In principle, privacy preservation techniques always deal with a trade-off between data utility and data protection, and they are supposed to balance these aims.


\begin{figure}[bt]
	\centering
	\includegraphics[width=0.88\textwidth]{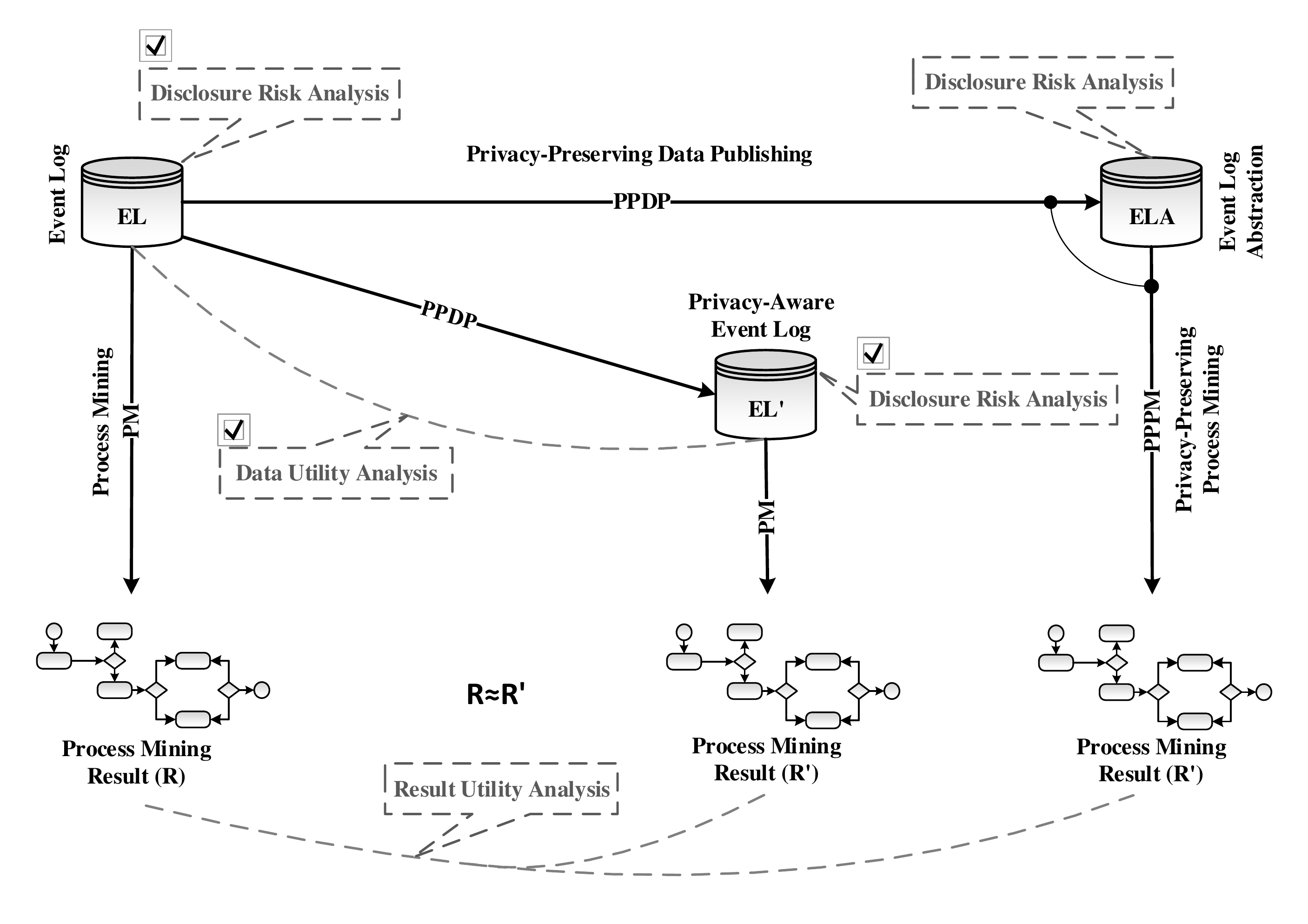}
	\caption{Overview of privacy-related activities in process mining. Privacy preservation techniques are applied to event logs to provide desired privacy requirements. The aim is to protect sensitive personal data, yet, at the same time, preserve data utility, and generate as similar as possible results to the original ones. The parts indicated by dashed callouts show the analyses that need to be performed to evaluate the effectiveness of privacy preservation techniques.}\label{fig:general_view}
\end{figure}

Figure~\ref{fig:general_view} shows the general view of privacy in process mining including two main activities: \textit{Privacy-Preserving Data Publishing} (PPDP) and \textit{Privacy-Preserving Process Mining} (PPPM). PPDP aims to hide the identity and the sensitive data of record owners in event logs to protect their privacy. PPPM aims to extend traditional process mining algorithms to work with the non-standard event data so-called \textit{Event Log Abstraction} (ELA) \cite{rafieippdp_short} that might result from PPDP techniques. \textit{Abstractions} are intermediate results, e.g., a directly follows graph could be an intermediate result of a process discovery algorithm. Note that PPPM algorithms are tightly coupled with the corresponding PPDP techniques. 

In this paper, our main focus is on the analyses indicated by the check-boxes in Fig.~\ref{fig:general_view}. Note that \textit{disclosure risk analysis} is done for a single event log, while for \textit{data/result utility analysis}, the original event log/result need to be compared with the privacy-aware event log/result. We consider simple event logs containing basic information for performing two main process mining activities: \textit{process discovery} and \textit{conformance checking}. We introduce two measures for quantifying disclosure risks in a simple event log: \textit{identity (case) disclosure} and \textit{attribute (trace) disclosure}. 
Using these measures, we show that even simple event logs could disclose sensitive information. We also propose a measure for quantifying \textit{data utility} which is based on the \textit{earth mover's distance}. 
So far, the proposed privacy preservation techniques in process mining use the \textit{result utility} approach to demonstrate the utility preservation aspect which is not as precise and general as the \textit{data utility} approach, since it is highly dependent on the underlying algorithms. 
We advocate the proposed measures by assessing their functionality for quantifying the disclosure risks and data utility on real-life event logs before and after applying a privacy preservation technique with different parameters.

The remainder of the paper is organized as follows. Section~\ref{sec:related_work} outlines related work. In Section~\ref{sec:prelimineries}, formal models for event logs are presented. We explain the measures in Section~\ref{sec:privacyQuantification}. The experiments are described in Section~\ref{sec:experiments}, and Section~\ref{sec:conclusion} concludes the paper.

\section{Related Work}\label{sec:related_work}
In process mining, the research field of confidentiality and privacy is growing in importance. In \cite{van2016responsible_short}, \textit{Responsible Process Mining} (RPM) is introduced as the sub-discipline focusing on possible negative side-effects of applying process mining. 
In \cite{MichaelKMBR19_short}, the authors propose a privacy-preserving system design for process mining, where a user-centered view is considered to track personal data. 
In \cite{rafieiWA19_short}, a framework is introduced providing a generic scheme for confidentiality in process mining. 
In \cite{rafiei2019role_short}, the authors introduce a privacy-preserving method for discovering roles from event data. 
In \cite{pretsaICPM2019_short}, the authors apply $k$-anonymity and $t$-closeness on event data to preserve the privacy of \textit{resources}. 
In \cite{MannhardtKBWM19_short}, the notion of \textit{differential privacy} is employed to preserve the privacy of \textit{cases}.
In \cite{rafieitlkc_short}, the $\tlkc$-privacy model is introduced to deal with high variability issues in event logs for applying group-based anonymization techniques.
In \cite{smcProcessMining_short}, a secure multi-party computation solution is proposed for preserving privacy in an inter-organizational setting.
In \cite{pika2020privacy_short}, the authors analyze data privacy and utility requirements for healthcare event data, and the suitability of privacy-preserving techniques is assessed. 
In \cite{rafieippdp_short}, privacy metadata in process mining are discussed and a privacy extension for the XES standard (https://xes-standard.org/) is proposed.

Most related to our work is \cite{riskProcessMining_short}, where a uniqueness-based measure is proposed to evaluate the re-identification risk of event logs. Privacy quantification in data mining is a well-developed field where the effectiveness of privacy preservation techniques is evaluated from different aspects such as \textit{dissimilarity} \cite{dissimilarity_short}, \textit{information loss} \cite{infoloss}, \textit{discernibility} \cite{discernibility_short}, and etc. We utilize the experiences achieved in this field and propose a trade-off approach as suggested in \cite{privacyQuantification_short}.


\section{Preliminaries}\label{sec:prelimineries}

In this section, we provide formal definitions for event logs used in the remainder. An event log is a collection of events, composed of different attributes, such that they are uniquely identifiable. In this paper, we consider only the mandatory attributes of events including \textit{case identifier}, \textit{activity name}, and \textit{timestamp}. Accordingly, we define a simple event, trace, and event log. In the following, we introduce some basic concepts and notations.

Let $A$ be a set. $A^*$ is the set of all finite sequences over $A$, and $\multiset(A)$ is the set of all multisets over the set $A$. 
For $A_1,A_2 \in \multiset(A)$, $A_1 \subseteq A_2$ if for all $a \in A$, $A_1(a) \leq A_2(a)$.
A finite sequence over $A$ of length $n$ is a mapping $\sigma \in \{1,...,n\} \rightarrow{A}$, represented as $\sigma = \langle a_1,a_2,...,a_n \rangle$ where $\sigma_i = a_i = \sigma(i)$ for any $1\leq i \leq n$, and $|\sigma|=n$. $a \in \sigma \Leftrightarrow{a=a_i}$ for $1 \leq i \leq n$. 
For $\sigma_1, \sigma_2 \in A^*$, $\sigma_1 \sqsubseteq \sigma_2$ if $\sigma_1$ is a subsequence of $\sigma_2$, e.g., $\langle a,b,c,x \rangle \sqsubseteq \langle z,x,a,b,b,c,a,b,c,x \rangle$. 
For $\sigma \in A^*$, $\{a \in \sigma\}$ is the set of elements in $\sigma$, and $[a \in \sigma]$ is the multiset of elements in $\sigma$, e.g., $[a \in \langle x,y,z,x,y \rangle ] = [x^2,y^2,z]$.

\begin{definition}[Simple Event]
	\label{def:event}
	A simple event is a tuple $e = (c,a,t)$, where $c \in \pcases$ is the \textit{case identifier}, $a \in \pactivities$ is the activity associated to event $e$, and $t \in \ptimes$ is the timestamp of event $e$.
	$\pi_X(e)$ is the projection of event $e$ on the attribute from domain $X$, e.g., $\pi_{\pactivities}(e) = a$.
	We call $\xi = \pcases \times \pactivities \times \ptimes$ the event universe.
\end{definition}

\begin{definition}[Simple Trace]
	\label{def:trace}
	Let $\xi$ be the universe of events. A trace $\sigma=\langle e_1,e_2,...,e_n \rangle$ in an event log is a sequence of events, i.e., $\sigma \in \xi^*$, s.t., for each $e_i,e_j \in \sigma$: $\pi_{\pcases}(e_i)=\pi_{\pcases}(e_j)$, and $\pi_{\ptimes}(e_i) \le \pi_{\ptimes}(e_j)$ if $i < j$.
	A \textit{simple trace} is a trace where all the events are projected on the activity attribute, i.e., $\sigma \in \pactivities^*$.
\end{definition}

\begin{definition}[Simple Event Log]
	\label{def:eventlog}
	A \textit{simple event log} is a multiset of simple traces, i.e., $\eventlog \in \multiset(\pactivities^*)$. We assume each trace in an event log belongs to an individual and $\sigma \neq \langle\rangle$ if $\sigma \in L$.
	$A_\eventlog=\{ a \in \pactivities \mid \exists_{\sigma \in L}a \in \sigma \}$ is the set of activities in the event log $\eventlog$. 
	$\tilde{L}=\{ \sigma \in L \}$ is the set of unique traces (variants) in the event log $\eventlog$.
	We denote $\universe_{L}$ as the universe of event logs.
\end{definition}


\begin{definition}[Trace Frequency]
	\label{def:set_variants}
	Let $\eventlog$ be an event log, $f_L \in \tilde{L} \rightarrow{[0,1]}$ is a function which retrieves the relative frequency of a trace in the event log $\eventlog$, i.e., $f_L(\sigma)=\nicefrac{L(\sigma)}{|L|}$ and $\sum_{\sigma \in \tilde{L}}f_L(\sigma)=1$. 
\end{definition}

\begin{definition}[Event Log Entropy]
	\label{def:el_entropy}
	$ent \in \universe_{L} \rightarrow{\mathbb{R}_{\ge 0}}$ is a function which retrieves the entropy of traces in an event log, s.t., for $L \in \universe_{L}$, $ent(L)=-\sum_{\sigma \in \tilde{L}}f_L(\sigma)log_2{f_L(\sigma)}$. We denote $max\_ent(L)$ as the maximal entropy achieved when all the traces in the event log are unique, i.e., $|\tilde{L}| = |L|$.
\end{definition}

\section{Privacy Quantification}\label{sec:privacyQuantification}
We employ a \textit{risk-utility} model for quantifying privacy in process mining where \textit{disclosure risk} and \textit{utility loss} are measured to assess the effectiveness of privacy preservation techniques before and after applying the techniques.

\subsection{Disclosure Risk}\label{sec:risk}
In this subsection, we introduce \textit{identity/case disclosure} and \textit{attribute/trace disclosure} for quantifying disclosure risk of event logs. Identity disclosure quantifies how uniquely the trace owners, i.e., cases, can be re-identified. Attribute disclosure quantifies how confidently the sensitive attributes of cases (as individuals) can be specified.
As discussed in \cite{rafieitlkc_short}, traces play the role of both quasi-identifiers and sensitive attributes. That is, a complete sequence of activities, which belongs to a case, is sensitive person-specific information. At the same time, knowing a part of this sequence, as background knowledge, can be exploited to re-identify the trace owner.
In a simple event log, traces, i.e., sequence of activities, are the only available information. Therefore, \textit{attribute disclosure} can be seen as \textit{trace disclosure}.

In the following, we define \textit{set}, \textit{multiset}, and \textit{sequence} as three types of background knowledge based on traces in simple event logs that can be exploited for uniquely re-identifying the trace owners or certainly specifying their complete sequence of activities. 
Moreover, we consider a size for different types of background knowledge as their power, e.g, the \textit{set} background knowledge of size 3 is more powerful than the same type of background knowledge of size 2. Note that the assumed types of background knowledge are the most general ones, and more types can be explored. However, the general approach will be the same.

\begin{definition}[Background Knowledge 1 - Set]
	\label{def:bk_set}
	In this scenario, we assume that an adversary knows a subset of activities performed for the case, and this information can lead to the identity or attribute disclosure.
	Let $\eventlog$ be an event log, and $A_L$ be the set of activities in the event log $\eventlog$. We formalize this background knowledge by a function 
	$proj_{set}^{L} \in 2^{A_L} \rightarrow{2^{L}}$. For $A \subseteq A_L$, $proj_{set}^{L}(A) = [\sigma \in L \mid A \subseteq \{a \in \sigma \} ]$.
	We denote $cand_{set}^l(L)=\{ A \subseteq A_L \mid |A|=l \wedge proj^L_{set}(A) \neq [] \}$ as the set of all subsets over the set $A_L$ of size $l$ for which there exists matching traces in the event log.
\end{definition}


\begin{definition}[Background Knowledge 2 - Multiset]
	\label{def:bk_mult}
	In this scenario, we assume that an adversary knows a sub-multiset of activities performed for the case, and this information can lead to the identity or attribute disclosure.
	Let $\eventlog$ be an event log, and $A_L$ be the set of activities in the event log $\eventlog$. We formalize this background knowledge by a function 
	$proj_{mult}^{L} \in \multiset(A_L) \rightarrow{2^{L}}$. For $A \in  \multiset(A_L)$, $proj_{mult}^{L}(A) = [\sigma \in L \mid A \subseteq [a \in \sigma]]$.
	We denote $cand_{mult}^l(L)=\{ A \in \multiset(A_L) \mid |A|=l \wedge proj^L_{mult}(A) \neq [] \}$ as the set of all sub-multisets over the set $A_L$ of size $l$ for which there exists matching traces in the event log.
\end{definition}


\begin{definition}[Background Knowledge 3 - Sequence]
	\label{def:bk_seq}
	In this scenario, we assume that an adversary knows a subsequence of activities performed for the case, and this information can lead to the identity or attribute disclosure.
	Let $\eventlog$ be an event log, and $A_L$ be the set of activities in the event log $\eventlog$. We formalize this background knowledge by a function 
	$proj_{seq}^{L} \in A_L^* \rightarrow{2^{L}}$. For $\sigma \in A_L^*$, $proj_{seq}^{L}(\sigma) = [\sigma' \in L \mid \sigma \sqsubseteq \sigma']$.
	We denote $cand_{seq}^l(L)=\{ \sigma \in A_L^{*} \mid |\sigma|=l \wedge proj^L_{seq}(\sigma) \neq [] \}$ as the set of all subsequences of size (length) $l$, based on the activities in $A_L$, for which there exists matching traces in the event log.
\end{definition}


\begin{exmp}[background knowledge]\label{exmp:general}
	Let $L= [\langle a,b,c,d \rangle^{10}, \langle a,c,b,d \rangle^{20}, \langle a,d,\\b,d \rangle^{5}, \langle a,b,d,d \rangle^{15} ]$ be an event log. $A_L=\{a,b,c,d\}$ is the set of unique activities, and $cand_{set}^2(L)=\{ \{a,b\}, \{a,c\}, \{a,d\}, \{b,c\}, \{b,d\}, \{d,c\}\}$ is the set of candidates of the \textit{set} background knowledge of size $2$. For $A=\{b,d\} \in cand_{set}^2(L)$ as a candidate of the set background knowledge of size $2$, $proj_{set}^{L}(A) = [\langle a,b,c,d \rangle^{10}, \langle a,c,b,d \rangle^{20}, \langle a,d,b,d \rangle^{5}, \langle a,b,d,d \rangle^{15}]$. For $A=[b,d^2]$ as a candidate of the \textit{multiset} background knowledge, $proj_{mult}^{L}(A) = [\langle a,d,b,d \rangle^{5}, \langle a,b,d,d \rangle^{15}]$. Also, for $\sigma=\langle b,d,d \rangle$ as a candidate of the \textit{sequence} background knowledge, $proj_{seq}^{L}(\sigma) = [\langle a,b,d,d \rangle^{15}]$. 
\end{exmp}

As Example~\ref{exmp:general} shows, the strength of background knowledge from the weakest to the strongest w.r.t. the type is as follows: \textit{set}, \textit{multiset}, and \textit{sequence}, i.e., given the event log $\eventlog$, $proj^L_{seq}(\langle b,d,d \rangle) \subseteq proj^L_{mult}([b,d^2]) \subseteq proj^L_{set}(\{b,d\})$.

\subsubsection{Identity (Case) Disclosure}
\label{sec:case_disclosure}
We use the uniqueness of traces w.r.t. the background knowledge of size $l$ to measure the corresponding case disclosure risk in an event log. Let $\eventlog$ be an event log and $type \in \{set,mult,seq\}$ be the type of background knowledge. The case disclosure based on the background knowledge $type$ of size $l$ is calculated as follows:

\begin{equation}\label{eq:case_disclosure}
\small
cd_{type}^l(L)=\sum_{x \in cand_{type}^l(L)}\frac{\nicefrac{1}{|proj^L_{type}(x)|}}{|cand_{type}^l(L)|}
\end{equation}

Equation~\eqref{eq:case_disclosure} calculates the average uniqueness based on the candidates of background knowledge, i.e., $x \in cand_{type}^l(L)$. Note that we consider equal weights for the candidates of background knowledge. However, they can be weighted based on the various criteria, e.g., the sensitivity of the activities included. One can also consider the worst case, i.e., the maximal uniqueness, rather than the average value.

\begin{exmp}[insufficiency of case disclosure analysis]\label{exmp:case_disclosure}
	Consider $L_1=[\langle a,\\b,c,d \rangle, \langle a,c,b,d \rangle, \langle a,b,c,c,d \rangle, \langle a,b,b,c,d \rangle]$ and $L_2=[\langle a,b,c,d \rangle^{4},\langle e,f \rangle^{4},\langle g,h \rangle^{4}]$ as two event event logs. $A_{L_1}=\{a,b,c,d\}$ and $A_{L_2}=\{a,b,c,d,e,f,g,h\}$ are the set of unique activities in $L_1$ and $L_2$, respectively. $cand_{set}^1(L_1)= \{\{a\},\{b\},\{c\},\{d\\\}\}$ and $cand_{set}^1(L_2) = \{\{a\},\{b\},\{c\},\{d\},\{e\},\{f\},\{g\},\{h\}\}$ are the set of candidates of the \textit{set} background knowledge of size $1$. Both event logs have the same value as the case disclosure for the \textit{set} background knowledge of size $1$ ($cd_{set}^1(L_1)=cd_{set}^1(L_2)=1/4$). However, in $L_2$, the complete sequence of activities performed for a victim case is disclosed by knowing only one activity without uniquely identifying the corresponding trace.
\end{exmp}

Example~\ref{exmp:case_disclosure} clearly shows that measuring the uniqueness alone is insufficient to demonstrate disclosure risks in event logs and the uncertainty in the set of sensitive attributes matching with the assumed background knowledge need to be measured, as well.
In the following, we define a measure to quantify the uncertainty in the set of matching traces. Note that, the same approach can be exploited to quantify the disclosure risk of any other sensitive attribute matching with some background knowledge.

\subsubsection{Attribute (Trace) Disclosure}
\label{sec:trace_disclosure}
We use the entropy of matching traces w.r.t. background knowledge of size $l$ to measure the corresponding trace disclosure risk in an event log. 
Let $\eventlog$ be an event log and $type \in \{set,mult,seq\}$ be the type of background knowledge. The trace disclosure based on the background knowledge $type$ of size $l$ is calculated as follows:

\begin{equation}\label{eq:trace_disclosure}
\small
td_{type}^l(L){=}1-{\sum_{x \in cand_{type}^l(L)}{\frac{\nicefrac{ent(proj^L_{type}(x))}{max\_ent(proj^L_{type}(x))}}{|cand_{type}^l(L)|}}}
\end{equation}

In \eqref{eq:trace_disclosure}, $max\_ent(proj^L_{type}(x))$ is the maximal entropy for the matching traces based on the type and size of background knowledge, i.e., uniform distribution of the matching traces. As discussed for \eqref{eq:case_disclosure}, in \eqref{eq:trace_disclosure}, we also assume equal weights for the candidates of background knowledge. However, one can consider different weights for the candidates. Also, the worst case, i.e., the minimal entropy, rather than the average entropy can be considered. 

The trace disclosure of the event logs in Example~\ref{exmp:case_disclosure} is as follows: $td_{set}^1(L_1)=0$ (the multiset of matching traces has the maximal entropy) and $td_{set}^1(L_2)=1$ (the entropy of matching traces is 0). These results distinguish the disclosure risk of the event logs.

\subsection{Utility Loss}\label{sec:utility}
In this subsection, we introduce a measure based on the \textit{earth mover's distance} \cite{emd} for quantifying the utility loss after applying a privacy preservation technique to an event log. The \textit{earth mover's distance} describes the distance between two distributions. In an analogy, given two piles of earth, it expresses the effort required to transform one pile into the other.
First, we introduce the concept of reallocation indicating how an event log is transformed into another event log. Then, we define a trace distance function expressing the cost of transforming one trace into another one. Finally, we introduce the utility loss measure that indicates the entire cost of transforming an event log to another one using the introduced reallocation and distance functions.

\subsubsection{Reallocation}
Let $\eventlog$ be the original event log and $L'$ be an anonymized event log derived from the original event log. 
We introduce $r \in \tilde{L} \times \tilde{L}' \rightarrow{[0,1]}$ as a function that indicates the movement of frequency between two event logs.
$r(\sigma,\sigma')$ describes the relative frequency of $\sigma \in \tilde{L}$ that should be transformed to $\sigma' \in \tilde{L}'$.
To make sure that a reallocation function properly transforms $\eventlog$ into $L'$, the frequency of each $\sigma \in \tilde{L}$ should be considered, i.e., for all $\sigma \in \tilde{L}$, $f_L(\sigma)= \sum_{\sigma' \in \tilde{L}'}r(\sigma, \sigma')$.
Similarly, the probability mass of traces $\sigma' \in \tilde{L}'$ should be preserved, i.e., for all $\sigma' \in \tilde{L}'$, $f_{L'}(\sigma')= \sum_{\sigma \in \tilde{L}}r(\sigma,\sigma')$. We denote $\mathcal{R}$ as the set of all reallocation functions which depends on $\eventlog$ and $L'$.

\begin{table}[tb]
	\centering
	\scriptsize
	\caption{The dissimilarity between two event logs based on the earth mover's distance assuming $r_s$ as a reallocation function and $d_s$ as the normalized Levenshtein distance.}\label{tbl:emd}
	\begin{tabular}{|l|c|c|c|c|}
		\hline
		\multicolumn{1}{|c|}{$r_s \cdot d_s$} & $\langle a,b,c,d  \rangle$ & $\langle a,c,b,d \rangle$ & $\langle a,e,c,d \rangle^{49}$ & $\langle a,e,b,d \rangle^{49}$ \\ \hline
		$\langle a,b,c,d \rangle^{50}$                       & $0.01 \cdot 0$            & $0 \cdot 0.5$              & $0.49 \cdot 0.25$        & $0 \cdot 0.5$              \\ \hline
		$\langle a,c,b,d \rangle^{50}$                       & $0 \cdot 0.5$               & $0.01 \cdot 0$           & $0 \cdot 0.5 $             & $0.49 \cdot 0.25$        \\ \hline
	\end{tabular}
\end{table}

\subsubsection{Trace Distance}
A trace distance function $d \in \pactivities^* \times \pactivities^* \rightarrow{[0,1]}$ expresses the distance between traces. 
This function is 0 if and only if two traces
are equal, i.e., $d(\sigma,\sigma')=0{\iff}\sigma=\sigma'$.
This function should also be symmetrical, i.e., $d(\sigma,\sigma')=d(\sigma',\sigma)$. 
Different distance functions can be considered satisfying these conditions. We use the \textit{normalized string edit distance} (Levenshtein) \cite{levenstein}.

\subsubsection{Utility Loss}
Let $\eventlog$ be an original event log, and $L'$ be an anonymized event log derived from the original event log. Several reallocation functions might exist.
However, the \textit{earth mover's distance} problem aims to express the shortest distance between the two event logs, i.e., the least mass movement
over the least distance between traces.
Therefore, the difference between $\eventlog$ and $L'$ using a reallocation function $r$ is the inner product of reallocation and distance. The data utility preservation is considered as $du(L,L') = 1 - \min\limits_{r \in \mathcal{R}} ul(r,L,L')$.

\begin{equation}\label{eq:utility_loss}
\small
ul(r,L,L')=r \cdot d =  \sum_{\sigma \in \tilde{L}} \sum_{\sigma' \in \tilde{L}'} {r(\sigma,\sigma')d(\sigma,\sigma')}
\end{equation}

\begin{exmp}\label{exmp:emd}\textnormal{\textbf{(using earth mover's distance to calculate dissimilarity between event logs)}}
	Let $L=[\langle a,b,c,d \rangle, \langle a,c,b,d \rangle, \langle a,e,c,d \rangle^{49}, \langle a,e,b,d \rangle^{49}]$ and $L'=[\langle a,b,c,d \rangle^{50},\langle a,c,b,d \rangle^{50}]$ be the original and aninymized event logs, respectively. Table~\ref{tbl:emd} shows the calculations assuming $r_s$ as a reallocation function and $d_s$ as the normalized Levenshtein distance, e.g., $r_s(\langle a,b,c,d \rangle,\langle a,e,c,d \rangle) = 0.49$ and $d_s(\langle a,b,c,d \rangle,\langle a,e,c,d \rangle)=0.25$. $ul(r_s,L,L')=0.24$ and $du(L,L')=0.76$. 
\end{exmp}

\begin{table}[b]
	\centering
	\vspace{-5 mm}
	\scriptsize
	\caption{The general statistics of the event logs used in the experiments.}\label{tbl:general_statistics}
	\begin{tabular}{|l|l|l|c|c|c|c|c|}
		\hline
		\multicolumn{3}{|c|}{\textbf{Event Log}} & \#traces & \#variants & \#events & \begin{tabular}[c]{@{}c@{}}\#unique\_activities\end{tabular} & \begin{tabular}[c]{@{}c@{}}trace\_uniqueness\end{tabular} \\ \hline
		\multicolumn{3}{|l|}{Sepsis-Cases \cite{Sepcis_2016_Felix_short}}       & 1050     & 845        & 15214    & 16                                                              & 80\%                                                         \\ \hline
		\multicolumn{3}{|l|}{BPIC-2017-APP \cite{BPIC2017_short}}      & 31509    & 102        & 239595   & 10                                                              & 0.3\%                                                        \\ \hline
	\end{tabular}
\end{table}

\section{Experiments}\label{sec:experiments}
In this section, we demonstrate the experiments on real-life event logs to advocate the proposed measures. We employ two human-centered event logs, where the \textit{case identifiers} refer to individuals. Sepsis-Cases \cite{Sepcis_2016_Felix_short} is a real-life event log containing events of sepsis cases from a hospital. BPIC-2017-APP \cite{BPIC2017_short} is also a real-life event log pertaining to a loan application process of a Dutch financial institute. We choose these event logs because they are totally different w.r.t. the uniqueness of traces. Table~\ref{tbl:general_statistics} shows the general statistics of these event logs. Note that \textit{variants} are the unique traces, and $trace\_uniquness = \nicefrac{\#variants}{\#traces}$. The implementation as a Python program is available on GitHub.\footnote{https://github.com/m4jidRafiei/privacy\_quantification}

\subsection{Disclosure Risk Analysis}\label{sec:risk_exp}
\begin{figure*}[t]
	\centering
	\subfloat[\scriptsize Sepsis-Cases \cite{Sepcis_2016_Felix_short}.]{\includegraphics[width=0.50\textwidth]{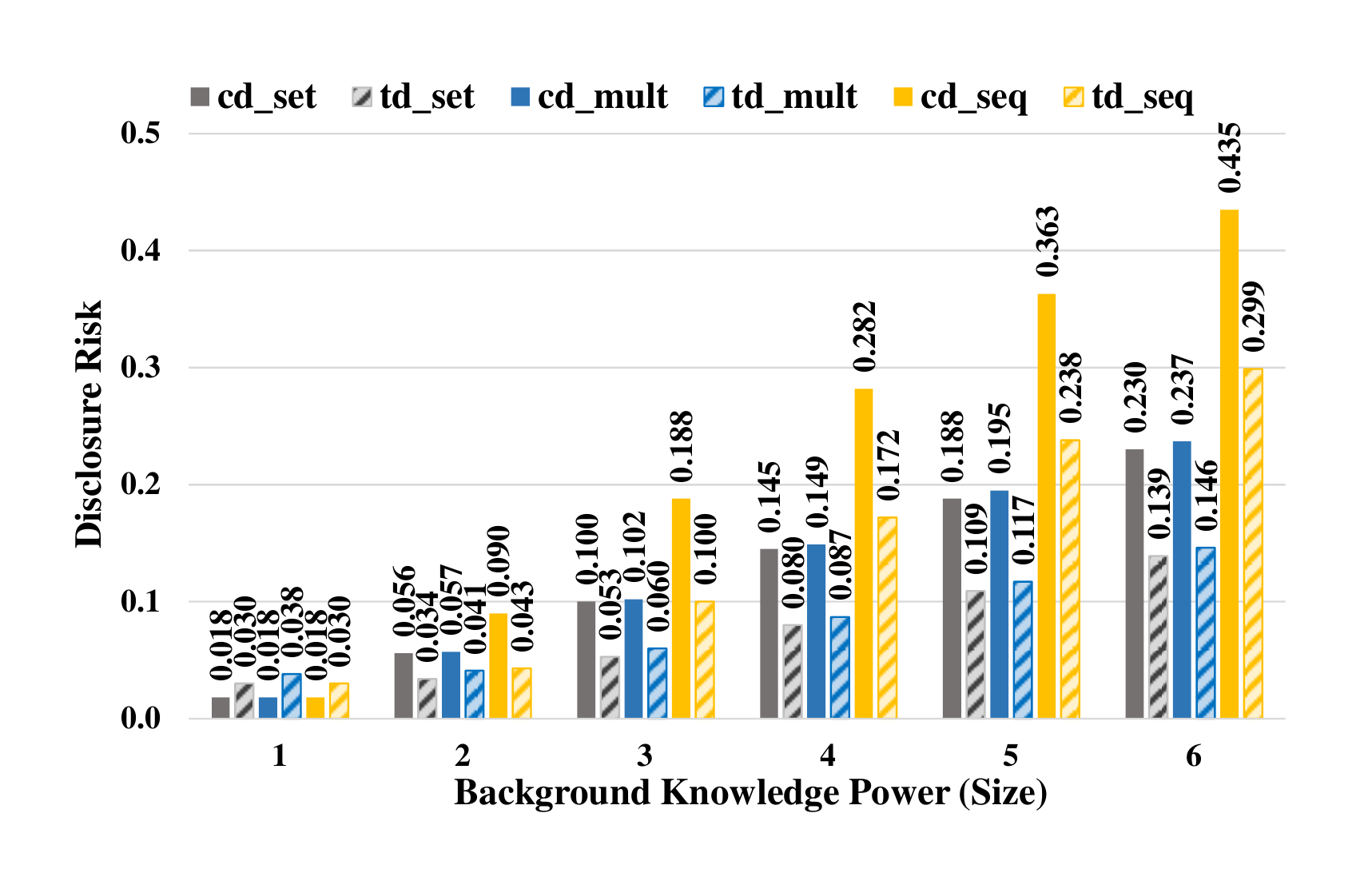}\label{fig:disclosure_sepsis}}
	\hfill
	\subfloat[\scriptsize BPIC-2017-APP \cite{BPIC2017_short}.]{\includegraphics[width=0.50\textwidth]{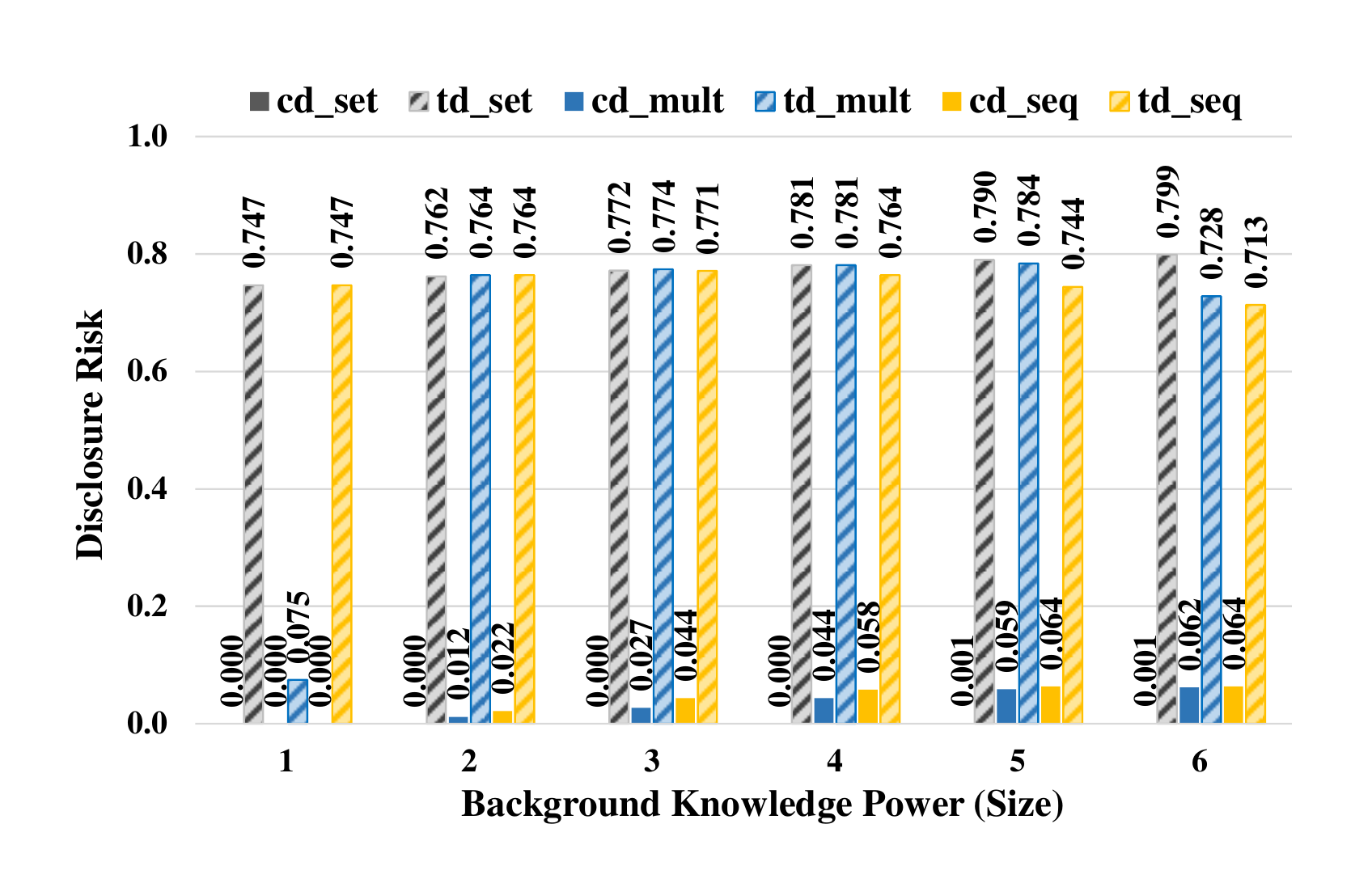}\label{fig:disclosure_BPIC}}
	\caption{Analyses of the case disclosure ($cd$) and the trace disclosure ($td$) based on the three types of background knowledge (i.e., $set$, $mult$, and $seq$) when we vary the background knowledge power (size) from 1 to 6. For example, in the Sepsis-Cases event log, the case disclosure risk of the background knowledge $seq$ ($cd\_seq$) of size 3 is 0.188.}
	\label{fig:disclosure}
\end{figure*}

In this subsection, we show the functionality of the proposed measures for disclosure risk analysis. To this end, we consider three types of background knowledge (\textit{set}, \textit{multiset}, and \textit{sequence}) and vary the background knowledge power (size) from 1 to 6. 
Figure~\ref{fig:disclosure_sepsis} shows the results for the Sepsis-Cases event log where the uniqueness of traces is high. As shown, the disclosure risks are higher for the more powerful background knowledge w.r.t. the \textit{type} and \textit{size}.

Figure~\ref{fig:disclosure_BPIC} demonstrates the results for the BPIC-2017-APP event log, where the uniqueness of traces is low. As shown, the case disclosure risk is low, which is expected regarding the low uniqueness of traces. However, the trace disclosure risk is high which indicates low entropy (uncertainty) of the traces. Moreover, for the stronger background knowledge w.r.t. the size, one can assume a higher case disclosure risk. However, the trace disclosure risk is correlated with the entropy of the sensitive attribute values and can be a high value even for weak background knowledge. 
The above-mentioned analyses clearly show that uniqueness alone cannot reflect the actual disclosure risk in an event log.

\subsection{Utility Loss Analysis}\label{sec:utility_exp}
In this subsection, we demonstrate the functionality of the proposed measure in Section~\ref{sec:utility} for quantifying data utility preservation after applying a privacy preservation technique. We use PPDP-PM \cite{rafieippdpTool_short} as a privacy preservation tool for process mining to apply the $\tlkc$-privacy model \cite{rafieitlkc_short} to a given event log. The $\tlkc$-privacy model is a group-based privacy preservation technique which provides a good level of flexibility through various parameters such as the type and size (power) of background knowledge. The $\mathrm{T}$ in this model refers to the accuracy of timestamps in the privacy-aware event log, $\mathrm{L}$ refers to the power of background knowledge\footnote{Note that this $\mathrm{L}$ is identical to the $l$ introduced as the power (size) of background knowledge and should not be confused with $\eventlog$ as the event log notation.}, $\mathrm{K}$ refers to the $k$ in the $k$-anonymity definition \cite{sweeney2002k}, and $\mathrm{C}$ refers to the bound of confidence regarding the sensitive attribute values in an equivalence class.

\begin{figure*}[t]
	\centering
	\subfloat[\scriptsize Using \textit{set} as background knowledge.]{\includegraphics[width=0.50\textwidth]{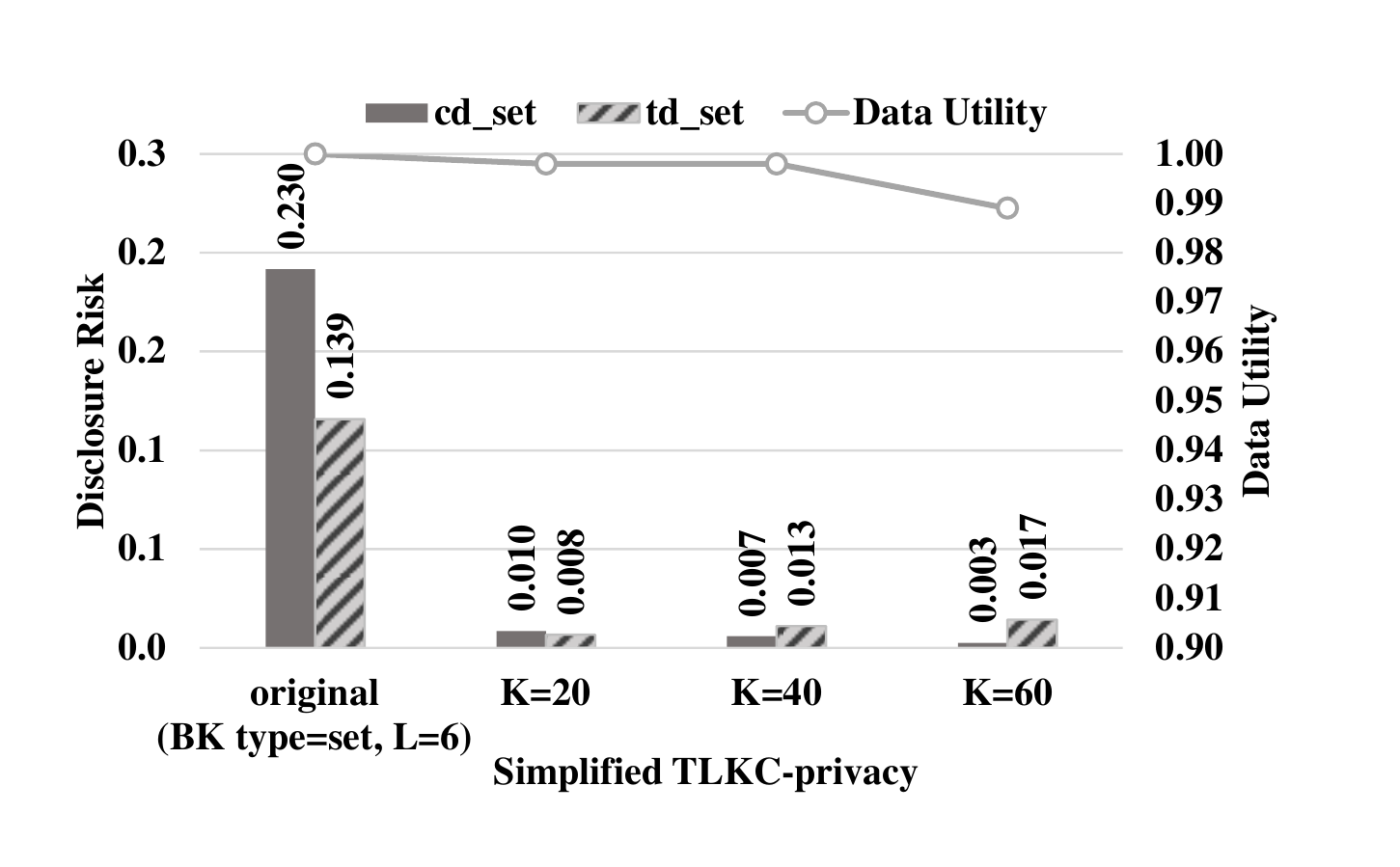}\label{fig:utility_set}}
	\hfill
	\subfloat[\scriptsize Using \textit{sequence} as background knowledge.]{\includegraphics[width=0.50\textwidth]{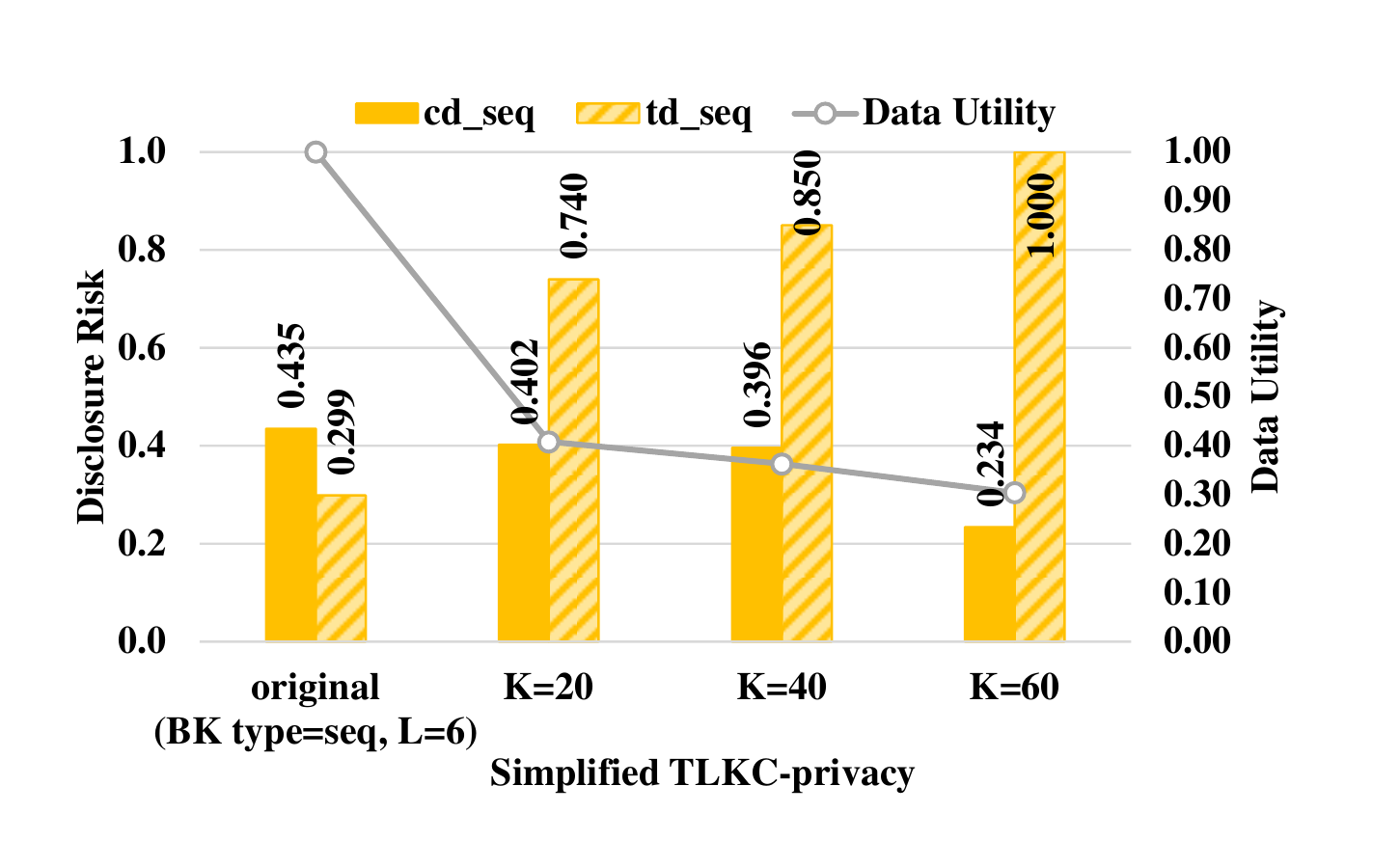}\label{fig:utility_seq}}
	\caption{The utility loss and disclosure risk analyses for the Sepsis-Cases event log where the background knowledge is \textit{set} or \textit{sequence}, and the power (size) of background knowledge is 6.}
	\label{fig:utility}
\end{figure*}

\begin{table*}[pb]
	\centering
	\scriptsize
	\caption{The general statistics before and after applying the $\tlkc$-privacy model.}\label{tbl:statistics_sepsis}
	\begin{tabular}{|c|c|c|c|c|c|c|}
		\hline
		\multicolumn{3}{|c|}{Event Log}                                                                                                       & \#traces & \#variants & \#events & \#unique\_activities \\ \hline
		\multicolumn{3}{|l|}{Original Sepsis-Cases}                                                                                           & 1050     & 845        & 15214    & 16                  \\ \hline
		\multirow{6}{*}{Anonymized Sepsis-Cases} & \multirow{3}{*}{\begin{tabular}[c]{@{}c@{}}BK type=set \\ BK size (L)=6\end{tabular}} & K=20 & 1050     & 842        & 15103    & 12                  \\ \cline{3-7} 
		&                                                                                     & K=40 & 1050     & 842        & 14986    & 11                  \\ \cline{3-7} 
		&                                                                                     & K=60 & 1050     & 818        & 14809    & 11                  \\ \cline{2-7} 
		& \multirow{3}{*}{\begin{tabular}[c]{@{}c@{}}BK type=seq\\ BK size (L)=6\end{tabular}}  & K=20 & 1050     & 34         & 3997     & 6                   \\ \cline{3-7} 
		&                                                                                     & K=40 & 1050     & 33         & 4460     & 5                   \\ \cline{3-7} 
		&                                                                                     & K=60 & 1050     & 18         & 3448     & 4                   \\ \hline
	\end{tabular}
\end{table*}

Assuming \textit{set} (\autoref{def:bk_set}) and \textit{sequence} (\autoref{def:bk_seq}) as the types of background knowledge, we apply the $\tlkc$-privacy model to the Sepsis-Cases event log with the following parameters: $\mathrm{L}=6$ (as the maximum background knowledge power in our experiments), $\mathrm{K}=\{20,40,60\}$, $\mathrm{C}=1$ (there is no additional sensitive attribute in a simple event log), and $\mathrm{T}$ is set to the maximal precision ($\mathrm{T}$ has no effect on a simple event log). That is, the $\tlkc$-privacy model is simplified to $k$-anonymity where the \textit{quasi-identifier} (background knowledge) is the \textit{set} or \textit{sequence} of activities. 
Table~\ref{tbl:statistics_sepsis} demonstrates the general statistics of the event logs before and after applying the privacy preservation technique.

%
%

Figure~\ref{fig:utility_set} shows disclosure risk and data utility analyses for the background knowledge \textit{set}, and Fig.~\ref{fig:utility_seq} shows the same analyses for the background knowledge \textit{sequence}. In both types of background knowledge, the data utility value decreases. For the stronger background knowledge, i.e., \textit{sequence}, the utility loss is much higher which is expected w.r.t. the general statistics in Table~\ref{tbl:statistics_sepsis}. However, the data utility for the weaker background knowledge remains high which again complies with the general statistics.
Note that since we apply $k$-anonymity (simplified $\tlkc$-model) only \textit{case disclosure}, which is based on the uniqueness of traces, decreases. Moreover, for the \textit{sequence} background knowledge, higher values for $\mathrm{K}$ result in more similar traces. Therefore, the \textit{trace disclosure} risk, in the anonymized event logs, drastically increases. These analyses demonstrate that privacy preservation techniques should consider different aspects of disclosure risk while balancing data utility preservation and sensitive data protection.

\section{Conclusion}\label{sec:conclusion}
Event logs often contain highly sensitive information, and regarding the rules imposed by regulations, these sensitive data should be analyzed responsibly.
Therefore, privacy preservation in process mining is recently receiving more attention. Consequently, new measures need to be defined to evaluate the effectiveness of the privacy preservation techniques both from the sensitive data protection and data utility preservation point of views. 
In this paper, using a trade-off approach, we introduced two measures for quantifying disclosure risks: \textit{identity/case disclosure} and \textit{attribute/trace disclosure}, and one measure for quantifying \textit{utility loss}. The introduced measures were applied to two real-life event logs. We showed that even simple event logs could reveal sensitive information. Moreover, for the first time, the effect of applying a privacy preservation technique on \textit{data utility} rather than \textit{result utility} was explored. The \textit{data utility} measure is based on the \textit{earth mover's distance} and can be extended to evaluate the utility w.r.t. the different perspectives of process mining, e.g., \textit{time}, \textit{resource}, etc.

\section*{Acknowledgment} Funded under the Excellence Strategy of the Federal Government and the L{\"a}nder. We also thank the Alexander von Humboldt (AvH) Stiftung for supporting our research.

\bibliographystyle{splncs04}
\bibliography{Refrences}

\end{document}